\documentclass[12pt]{article}
\usepackage{algpseudocode}
\usepackage{amsmath, amssymb, amsthm}
\usepackage{float, fullpage, graphicx, multirow,parskip, subcaption, setspace}
\usepackage{comment}
\usepackage{url}
\usepackage{enumitem}
\usepackage{tikz}
\usepackage{bm, bbm}
\usetikzlibrary{shapes.geometric, positioning}
\usetikzlibrary{quotes, angles}
\usepackage{rotating}
\usepackage{hyperref}
\usepackage[capitalize]{cleveref}
\usepackage{natbib}
\usepackage{placeins}
\usepackage{algorithm2e}
\RestyleAlgo{ruled}

\definecolor{SkyBlue}{RGB}{14, 118, 188}
\definecolor{BrightRed}{RGB}{223,82, 78}
\newcommand{\iidsim}{\stackrel{iid}{\sim}}
\newcommand{\indsim}{\stackrel{ind}{\sim}}

\hypersetup{pdfborder = {0 0 0.5 [3 3]}, colorlinks = true, linkcolor = BrightRed, citecolor = SkyBlue}

\def\keywordname{{\bfseries \emph Keywords}}%
\def\keywords#1{\par\addvspace\medskipamount{\rightskip=0pt plus1cm
\def\and{\ifhmode\unskip\nobreak\fi\ $\cdot$
}\noindent\keywordname\enspace\ignorespaces#1\par}}

\title{Wild posteriors in the wild}
\author{Yunyi Shen\thanks{EECS, MIT, yshen99@mit.edu} \and Tamara Broderick\thanks{EECS, MIT, tbroderick@mit.edu}}

\begin{document}
\def\bY{\bm{Y}}
\def\by{\bm{y}} 

\def\bz{\bm{z}}
\def\bX{\bm{X}}
\def\bx{\bm{x}} 

\def\R{\mathbb{R}}
\def\N{\mathcal{N}}
\def\P{\mathbb{P}}
\def\E{\mathbb{E}}

\def\Xcal{\mathcal{X}}

\maketitle
\begin{abstract}
Bayesian posterior approximation has become more accessible to practitioners than ever, thanks to modern black-box software. While these tools provide highly accurate approximations with minimal user effort, certain posterior geometries remain notoriously difficult for standard methods. As a result, research into alternative approximation techniques continues to flourish. In many papers, authors validate their new approaches by testing them on posterior shapes deemed challenging or ``wild.'' However, these shapes are not always directly linked to real-world applications where they naturally occur. In this note, we present examples of practical applications that give rise to some commonly used benchmark posterior shapes.
\end{abstract}

\section{Introduction}
Bayesian posterior approximation is more accessible to practitioners than ever thanks to modern black-box software --- such as Stan \citep{carpenter2017stan}, Pyro \citep{bingham2019pyro}, PyMC \citep{abril2023pymc}, NIMBLE \citep{de2017nimble}, and others. While this software offers widely accurate approximation with minimal user effort, it is well known that certain posterior geometries remain challenging for standard approximation schemes. As such, research into alternative approximations continues to thrive. In these papers, it is common for authors to demonstrate that their new approximation works well by testing it on posterior shapes considered to be challenging or ``wild.'' But the shapes are not always directly connected to a practical application where they might arise. In the present note, we provide examples of applications in the wild that give rise to some common benchmark posterior shapes. We hope these connections to applications will be useful for developers of posterior approximations in at least two ways. (1) Understanding the underlying application and model can help a developer understand precisely what the user hopes to get out of the data analysis. For instance, a posterior mean and variance need not always be useful posterior summaries. (2) In cases where a posterior shape can be matched to a modern data analysis, the developer can rest assured that a good approximation will be useful for applied problems. While the present note cannot be exhaustive, we collect further examples at 
\url{https://github.com/YunyiShen/weird-posteriors}.
And we hope our work inspires developers to track down applications corresponding to their benchmark shapes.

\section{Wild posteriors in the wild}
\begin{figure}
    \centering
    \includegraphics[width=\linewidth]{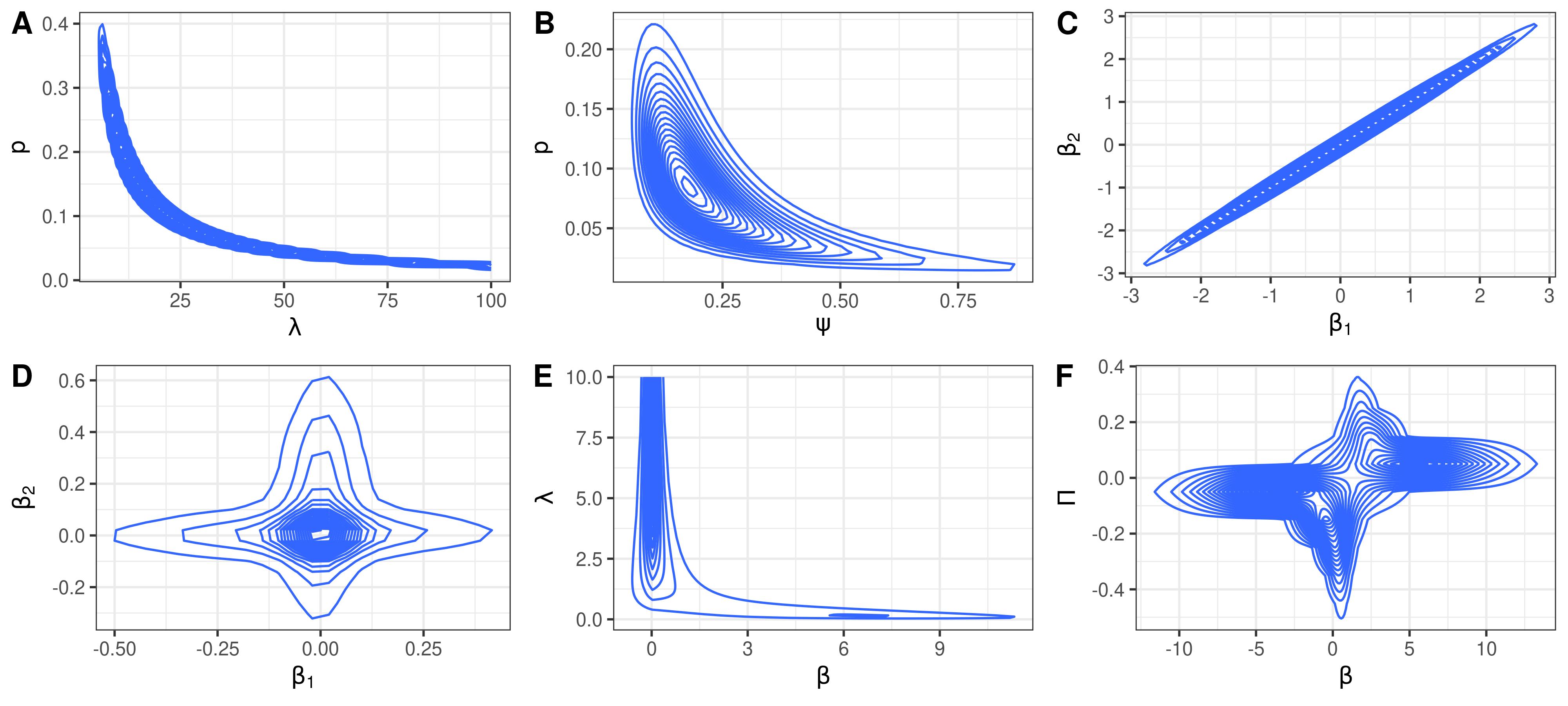}
    \caption{Example unnormalized posterior contour. A: N-mixture model, B: Occupancy model, C: linear regression with collinearity, D: Spike-and-slab prior, E: adaptive LASSO, F: weak instrumental variable. }
    \label{fig:samples}
\end{figure}

\textbf{Banana.} The contours of a posterior distribution can take a ``banana'' shape when data provides information about the product of two parameters but can only weakly identify the two. As one example, consider an \textbf{N-mixture} model \citep{royle2004n}, used by ecologists counting unmarked animals.
\begin{equation}
\begin{aligned}
    &\text{observing:}~~y_{i,n_i}\\
    &y_{i,n_i}\indsim \text{Binomial}(p, N_i),~~N_i\iidsim \text{Poisson}(\lambda)~~~~\text{(likelihood)}\\
    &p\sim \text{Unif}(0,1),~~\lambda\sim \text{Gamma}(\alpha, \beta)~~~~\text{(prior)}
\end{aligned}
\label{eq:nmixture}
\end{equation}
Here $y_{i,n_i}$ is the count of the animal at location $i$ and ``repeat'' $n_i$. The goal is to infer the ``abundance'' $\lambda$ and ``detection rate'' $p$. If for each location $i$ there are no repeats (i.e., $n_i=1$), then the data is Poisson distributed with parameter $\lambda p$, so $\lambda$ and $p$ are not identified. With some repeats at each location, these two parameters are weakly identified (\cref{fig:samples}-A). We give the details for all of our simulations in \cref{app:expdetails}.

A second example is the \textbf{occupancy} model \citep{mackenzie2002estimating}.
\begin{equation}
    \begin{aligned}
        &\text{observing:}~~y_{i,n_i}\\
        &y_{i, n_i}\indsim \text{Bernoulli}(pz_i),~~~~z_i\iidsim \text{Bernoulli}(\psi)~~~~\text{(likelihood)}\\
        &\psi, p\sim \text{Unif}(0,1)~~~~\text{(prior)}
    \end{aligned}
    \label{eq:occupancy}
\end{equation}
Here $y_{i, n_i}$ equals 1 if one sees an animal at location $i$ in repeat $n_i$ and otherwise $y_{i, n_i} = 0$. The goal is to infer the occupancy rate $\psi$ and the detection rate $p$. When most of the data observations are $0$, there are two competing explanations for the data: either small $p$ or small $\psi$, so the model is at best weakly identified (\cref{fig:samples}-B).

\textbf{Needle.}
A posterior can be ``needle''-shaped when the data provides information about the sum or difference of two parameters but can only weakly identify their values.
A familiar example is linear regression with \textbf{multicollinearity}. \Cref{fig:samples}-C shows an example with two linear regression coefficients.

\textbf{Cross.}
Consider a linear regression where the coefficients have a \textbf{spike-and-slab} Gaussian mixture prior \citep{george1993variable}. For instance, \citet{kazemi2024using} used this prior in genome-wide association studies to find variants related to bipolar disorder. When data is limited, the posterior can look like a ``cross'' (\cref{fig:samples}-D).
\begin{equation}
    \begin{aligned}
        &\text{observing:}~~(x_i, y_i)\\
        &y_i=x_i^\top \beta + \epsilon_i, \epsilon_i\iidsim \text{Normal}(0,1)~~~~\text{(likelihood)}\\
        &\beta_j\indsim \text{Normal}(0, 0.1z_j+100(1-z_j)),~~z_j\iidsim \text{Bernoulli}(0.1)~~~~\text{(prior)}    
    \end{aligned}
    \label{eq:spikeslab}
\end{equation}

\textbf{Multimodal.}
Multimodality is common in Bayesian variable selection. The Bayesian LASSO \citep{park2008bayesian} has a log-concave posterior and thus is not multimodal. However, the \textbf{adaptive LASSO}\footnote{This name appears in \citet{wang2012bayesianglasso}.} proposed in \citet{park2008bayesian} puts a hyperprior on the amount of penalization and can exhibit multimodality even with a single covariate (\cref{fig:samples}-E).
\begin{equation}
\begin{aligned}
    &\text{observing:}~~(x_i,y_{i}), \sigma\\
    &y_i\indsim N(x_i\beta,\sigma^2)~~~~\text{(likelihood)}\\
    &\beta\sim \text{Laplace}(0,\lambda)~~~~\lambda\sim \text{Unif}(0.001,10)~~~~\text{(prior)}
\end{aligned}
\label{eq:lasso}
\end{equation}
\citet{gallo2022mammals} use this prior (including the uniform hyperprior on $\lambda$ just above) to analyze how mammals adjust ``diel'' activity across a gradient of urbanization. \citet{van2019shrinkage} reviews the use of this style of prior in psychology, and \citet{banner2020use} critiques it in ecology.

\textbf{Singularity.}
\citet{hoogerheide2008possibly} showed that a simple \textbf{instrumental variable} model with a diffuse prior can exhibit challenging posterior behaviors, including point singularities, diverging ridgelines, and multiple modes. 
\citet{hoogerheide2008possibly} use this model to analyze the effect of education on income using birth quarter as an instrument (since, in the United States, birth date determines start and duration of schooling).
\begin{equation}
\begin{aligned}
    &\text{observing:}~~(x_i, y_i, z_i)\\
    &y_i=x_i\beta+\epsilon_i,~~x_i=z_i\Pi+v_i,~~(\epsilon_i,v_i)\iidsim \text{Normal}(0,\Sigma)~~~\text{(likelihood)}\\
    &p(\beta,\Pi, \Sigma)\propto |\Sigma|^{-3/2}~~~~\text{(prior)}
\end{aligned}
\label{eq:instrumental}
\end{equation}
One problem arises when $\Pi$ is close to 0; then the model is very weakly identified, and a singularity forms around $\Pi=0$ (\cref{fig:samples}-F).

\section{Discussion}
Bayesian posterior approximation methods continue to progress. To support the development of new methods, it is useful to have example models and datasets where challenging posteriors arise, complementing standard benchmarking distributions.

However, we note that a complex posterior shape often signals identifiability problems in the model. So in many cases, what may seem like a challenging posterior perhaps need not be. For example, in Gaussian mixture models, multimodality often arises since an ordering must be assigned (in code) to the elements of the fundamentally unordered partition of the data; then we obtain (redundant) modes for each potential labeling of the partition elements. In this case, capturing a single mode among these redundant modes would not only be sufficient, but would in fact be strictly more desirable than capturing the full posterior (cf.\ the well-known label-switching problem in Markov chain Monte Carlo samplers of mixtures \citep{jasra2005labelswitching}). In some cases, the practitioner might be able to choose an appropriate model that avoids unidentifiability or challenging shapes---such as using a non-centered parameterization for Neal's funnel \citep{neal2003slice}. But in other cases, such as the Gaussian mixture, the best approach might be to address the problem in the approximation software itself, perhaps by focusing on common types of data analyses.



\bibliography{references}

\begin{thebibliography}{}

\bibitem[Abril-Pla et~al., 2023]{abril2023pymc}
Abril-Pla, O., Andreani, V., Carroll, C., Dong, L., Fonnesbeck, C.~J.,
  Kochurov, M., Kumar, R., Lao, J., Luhmann, C.~C., Martin, O.~A., Osthege, M.,
  Vieira, R., Wiecki, T., and Zinkov, R. (2023).
\newblock {PyMC}: a modern, and comprehensive probabilistic programming
  framework in {Python}.
\newblock {\em PeerJ Computer Science}, 9:e1516.

\bibitem[Banner et~al., 2020]{banner2020use}
Banner, K.~M., Irvine, K.~M., and Rodhouse, T.~J. (2020).
\newblock The use of {Bayesian} priors in ecology: The good, the bad and the
  not great.
\newblock {\em Methods in Ecology and Evolution}, 11(8):882--889.

\bibitem[Bingham et~al., 2019]{bingham2019pyro}
Bingham, E., Chen, J.~P., Jankowiak, M., Obermeyer, F., Pradhan, N.,
  Karaletsos, T., Singh, R., Szerlip, P., Horsfall, P., and Goodman, N.~D.
  (2019).
\newblock Pyro: Deep universal probabilistic programming.
\newblock {\em Journal of Machine Learning Research}, 20(28):1--6.

\bibitem[Carpenter et~al., 2017]{carpenter2017stan}
Carpenter, B., Gelman, A., Hoffman, M.~D., Lee, D., Goodrich, B., Betancourt,
  M., Brubaker, M., Guo, J., Li, P., and Riddell, A. (2017).
\newblock Stan: A probabilistic programming language.
\newblock {\em Journal of Statistical Software}, 76:1--32.

\bibitem[de~Valpine et~al., 2017]{de2017nimble}
de~Valpine, P., Turek, D., Paciorek, C.~J., Anderson-Bergman, C., Lang, D.~T.,
  and Bodik, R. (2017).
\newblock Programming with models: writing statistical algorithms for general
  model structures with {NIMBLE}.
\newblock {\em Journal of Computational and Graphical Statistics},
  26(2):403--413.

\bibitem[Gallo et~al., 2022]{gallo2022mammals}
Gallo, T., Fidino, M., Gerber, B., Ahlers, A.~A., Angstmann, J.~L., Amaya, M.,
  Concilio, A.~L., Drake, D., Gay, D., Lehrer, E.~W., Murray, M.~H., Ryan,
  T.~J., St~Clair, C.~C., Salsbury, C.~M., Sander, H.~A., Stankowich, T.,
  Williamson, J., Belaire, J.~A., Simon, K., and Magle, S.~B. (2022).
\newblock Mammals adjust diel activity across gradients of urbanization.
\newblock {\em eLife}, 11:e74756.

\bibitem[George and McCulloch, 1993]{george1993variable}
George, E.~I. and McCulloch, R.~E. (1993).
\newblock Variable selection via {Gibbs} sampling.
\newblock {\em Journal of the American Statistical Association},
  88(423):881--889.

\bibitem[Hoogerheide and van Dijk, 2008]{hoogerheide2008possibly}
Hoogerheide, L.~F. and van Dijk, H.~K. (2008).
\newblock Possibly ill-behaved posteriors in econometric models.

\bibitem[Jasra et~al., 2005]{jasra2005labelswitching}
Jasra, A., Holmes, C.~C., and Stephens, D.~A. (2005).
\newblock Markov chain {Monte Carlo} methods and the label switching problem in
  bayesian mixture modeling.
\newblock {\em Statistical Science}, 20(1):50 -- 67.

\bibitem[Kazemi~Naeini et~al., 2024]{kazemi2024using}
Kazemi~Naeini, M., Akbarzadeh, M., Kazemi, I., Speed, D., and Hosseini, S.~M.
  (2024).
\newblock Using the {Bayesian} variational spike and slab model in a
  genome-wide association study for finding associated loci with bipolar
  disorder.
\newblock {\em Annals of Human Genetics}, 88(3):212--246.

\bibitem[Kim et~al., 1998]{kim1998stochastic}
Kim, S., Shephard, N., and Chib, S. (1998).
\newblock Stochastic volatility: likelihood inference and comparison with
  {ARCH} models.
\newblock {\em The Review of Economic Studies}, 65(3):361--393.

\bibitem[MacKenzie et~al., 2002]{mackenzie2002estimating}
MacKenzie, D.~I., Nichols, J.~D., Lachman, G.~B., Droege, S., Andrew~Royle, J.,
  and Langtimm, C.~A. (2002).
\newblock Estimating site occupancy rates when detection probabilities are less
  than one.
\newblock {\em Ecology}, 83(8):2248--2255.

\bibitem[Neal, 2003]{neal2003slice}
Neal, R.~M. (2003).
\newblock Slice sampling.
\newblock {\em The Annals of Statistics}, 31(3):705--767.

\bibitem[Park and Casella, 2008]{park2008bayesian}
Park, T. and Casella, G. (2008).
\newblock The {Bayesian} {LASSO}.
\newblock {\em Journal of the American Statistical Association},
  103(482):681--686.

\bibitem[Royle, 2004]{royle2004n}
Royle, J.~A. (2004).
\newblock N-mixture models for estimating population size from spatially
  replicated counts.
\newblock {\em Biometrics}, 60(1):108--115.

\bibitem[{Stan Development Team}, 2025]{stan:svm}
{Stan Development Team} (2025).
\newblock Stan reference guide: time-series models: Stochastic volatility
  models.
\newblock
  \url{https://mc-stan.org/docs/stan-users-guide/time-series.html\#stochastic-volatility-models},
  Accessed: 2025-02-24.

\bibitem[Van~Erp et~al., 2019]{van2019shrinkage}
Van~Erp, S., Oberski, D.~L., and Mulder, J. (2019).
\newblock Shrinkage priors for {Bayesian} penalized regression.
\newblock {\em Journal of Mathematical Psychology}, 89:31--50.

\bibitem[Wang, 2012]{wang2012bayesianglasso}
Wang, H. (2012).
\newblock Bayesian graphical {LASSO} models and efficient posterior
  computation.
\newblock {\em Bayesian Analysis}, 7(4):867 -- 886.

\bibitem[Wee, 2024]{wee2024comparing}
Wee, B. (2024).
\newblock Comparing {MCMC} algorithms in stochastic volatility models using
  simulation based calibration.
\newblock {\em arXiv preprint arXiv:2402.12384}.

\end{thebibliography}

\appendix
\section{Additional experiments}
\label{app:moreexp}
\textbf{Mushroom.}
The \textbf{stochastic volatility} model for stock returns proposed by \citet{kim1998stochastic} can give rise to a mushroom-shaped posterior. 
\begin{equation}
\begin{aligned}
    &\text{observing}: y_t\\
    &y_t= \epsilon_t e^{\frac{h_t}{2}},~~~~ h_{t+1} =\mu+\phi(h_t-\mu) +\delta_t\sigma\\
    &h_1\sim \text{Normal}\left( \mu, \frac{\sigma}{\sqrt{1-\phi^2}}\right),~~~~\epsilon_t, \delta_t\iidsim \text{Normal}(0,1)~~~\text{(likelihood)}\\
    &\phi\sim \text{Uniform}(-1,1),~~\sigma\sim \text{Cauchy}(0,5),~~\mu\sim \text{Cauchy}(0,10)~~~\text{(prior)}
\end{aligned}
\label{eq:mushroom}
\end{equation}
The challenge of this model arises from the posterior behavior when $\phi$ approaches 1. We next describe why the resulting shape might be seen as mushroom-like, why the mushroom shape arises, and finally why the mushroom shape is challenging.

\begin{figure}[htp]
    \centering
    \includegraphics[width=0.7\linewidth]{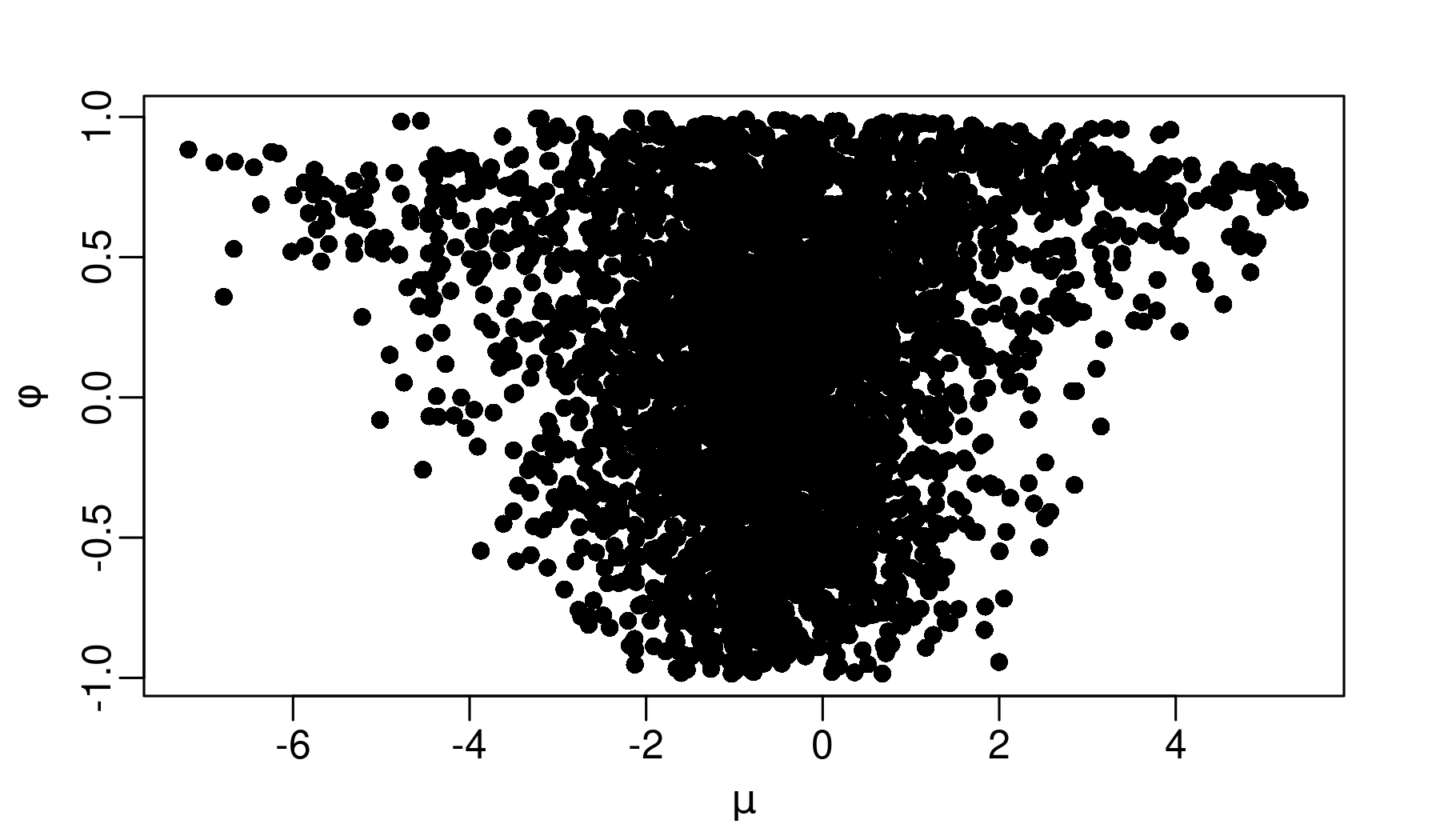}
    \caption{Posterior samples of the stochastic volatility model}
    \label{fig:stochvola}
\end{figure}

We first describe what the mushroom shape looks like. When $\phi$ is very near 1, $\mu$ is much heavier tailed than for $\phi$ substantially smaller than 1. When considering the marginal posterior over $\phi$ and $\mu$, we can think of the heavy-tailed behavior for $\phi$ near 1 as corresponding to the hat of the mushroom and the region with $\phi$ substantially smaller than 1 as corresponding to the stem of the mushroom. We provide a rough illustration in \cref{fig:stochvola}. For this model, it is difficult to access the unnormalized posterior marginal density in $\phi$ and $\mu$ due to the need to numerically integrate out all other parameters.
Therefore, to create \cref{fig:stochvola}, we show samples from Stan \citep{carpenter2017stan}. For $\phi$ near 1, we can see some indication of the widening behavior of $\mu$ in the plot.

Next we describe why the mushroom shape arises.
First, observe that we can rewrite the formula for $h_{t+1}$ as a function of $h_t$ as follows: $h_{t+1}=(1-\phi)\mu+\phi h_t+\delta_t \sigma$. As $\phi \rightarrow 1$, the first term vanishes. So $h_{t+1}$ depends on $\mu$ increasingly primarily through $h_{t}$ (rather than directly). By recursion, we conclude that, as $\phi \rightarrow 1$, $h_{t+1}$ comes to depend on $\mu$ primarily through $h_1$. Next, we observe that the dependency of $h_1$ on $\mu$ also becomes weaker as $\phi \rightarrow 1$. In particular, as $\phi \rightarrow 1$, the variance of the prior on $h_1$ diverges. So as $\phi \rightarrow 1$, there becomes increasingly little dependence of the data on $\mu$; that is to say, $\mu$ is increasingly weakly identified, and the posterior marginal over $\mu$ reverts to its (heavy-tailed) prior behavior.

Finally, we discuss why this shape is challenging. The resulting mushroom shape essentially leaves a thin slice (thin across $\phi$) is the marginal posterior over $\phi$ and $\mu$. Thus the stem and hat of the posterior exhibit fundamentally different length scales in terms of the size of the largest sphere that fits into posterior level sets. This challenge is the same one as the one faced by Neal's funnel \citep{neal2003slice}. Reparametrization might help address some of the challenges of sampling in this model \citep{wee2024comparing}. 


\section{Simulation details}
\label{app:expdetails}
We generate our contour plots in \cref{fig:samples} by plotting the unnormalized posteriors in each case.

To generate \cref{fig:samples}-A, we simulated data from the likelihood in \cref{eq:nmixture} with $\lambda=30$ and $p=0.1$. We generated data at 20 locations and 5 repeats at each location $i$ ($\forall i \in \{1,\ldots,20\}, N_i =5$).

To generate \cref{fig:samples}-B, we simulated data from the likelihood in \cref{eq:occupancy} with $\psi = p = 0.1$. We generated data at 100 locations and 8 repeats per location. $n_i = 1, \dots, 8$ for all $i$ ($\forall i \in \{1,\ldots,100\}, N_i =8$).

To generate \cref{fig:samples}-C, we use the following model where $\beta$ and $x_i$ are two-dimensional, respectively. 
\begin{equation}
\begin{aligned}
    &\text{observing:}~~(x_i,y_{i})\\
    &y_i\indsim N(x_i^\top\beta,1)~~~~\text{(likelihood)}\\
    &\beta\sim N(0,\sigma^2 I_2)~~~~\text{(prior)}
\end{aligned}
\end{equation}
Here, $I_2$ is the identity matrix of dimension 2.  Further, we take the $x_i$'s generated from a normal with high covariance; in particular, we simulate the $x_i$'s from a bivariate normal with variance components on the diagonal equal to 1 and off-diagonal covariance components equal to  $-0.995$. We choose $\beta = (-10,10)$. And we choose a large $\sigma$ (100) so that the prior is not very informative. Then the two $\beta$'s are only weakly identified up to their difference.

To generate \cref{fig:samples}-D, we simulated 10 data points with $\beta = (0,0)$ from the likelihood in \cref{eq:spikeslab}. We drew the $x_i$ values i.i.d.\ uniformly in $[-2,2]$.

To generate \cref{fig:samples}-E, we simulated we simulated 5 data points from the likelihood in \cref{eq:lasso} with $\beta=5$ and a known $\sigma=8$. We let all $x_i=1$, so the task was Gaussian mean estimation.

To generate \cref{fig:samples}-F, we simulated 50 data points from the likelihood in \cref{eq:instrumental} with $\beta = \Pi=0.1$, $\Sigma=I$, and instruments $z_i\iidsim \text{Bernoulli}(0.75)$.

To generate \cref{fig:stochvola}, we simulated from the model in \cref{eq:mushroom} with $\mu=-1.02,\phi=-0.95,\sigma=0.1$. We generated data for $t=1,\dots,5$. The implementation in Stan was taken from the Stan reference guide \citep{stan:svm}.

\end{document}